# Framework to model virtual factories: a digital twin view

Ali Ahmad Malik
Siemens Gamesa Renewable Energy, 7330 Brande, Denmark
Email: ali.malik@siemensgamesa.com
Mobile: +45 7158 22 21

## Pre-print

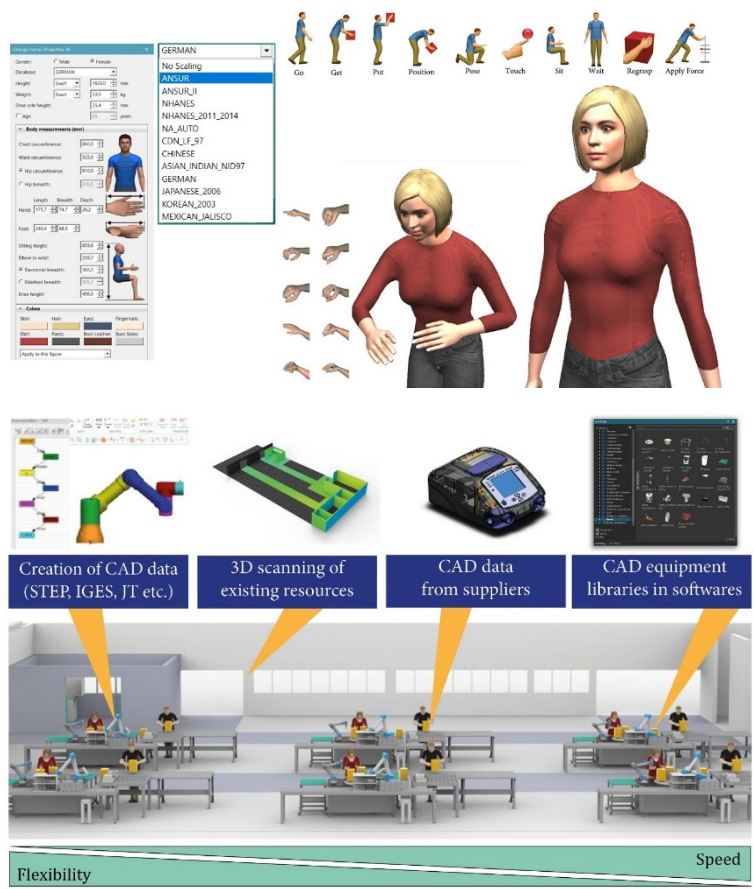

**Highlights:**

- Digital twin has emerged as the next wave in the simulation technology to answer the growing complexities of manufacturing systems.
- The exactitude of the virtual models defines the accuracy of predicted behavior obtained from the experimentation on the digital twin.
- The lack of frameworks often makes it difficult to achieve a model with accuracy, short lead time and with ability to be reconfigured.
- A VMDT framework for simulation modelling of a manufacturing system is presented.

# Framework to model virtual factories: a digital twin view

Ali Ahmad Malik

Siemens Gamesa Renewable Energy, 7330 Brande, Denmark

Email: ali.malik@siemensgamesa.com

Mobile: +45 7158 22 21

**Abstract:**

Digital twin has emerged as a technology to predict the 'undesirables' and ensure desired performance of complex systems. Although digital twins have got attention in the manufacturing research spectrum, yet their industrial application has been limited. Virtual simulations are considered an integral part of a digital twin, but a challenge is the lack of structured approaches of creating simulation models that can be extended as a digital twin. At the same time, the virtual models need to be accurate and flexible enough to be updated along the life cycle of the factory as desired in a digital twin. This paper presents a framework for virtual modeling of factories that can be extended as a digital twin. Case of a manufacturing company is presented to model and simulate a manufacturing system in a structured yet flexible way.

**Keywords:** Simulation; Digital twin; Manufacturing; Industry 4.0; Virtual.

## 1. Introduction

A digital twin is a data connected 3D (three dimensional) visualization of elements and dynamics of a physical system (Malik & Brem 2020). The vision of a digital twin (DT) is the integration of a virtual representation of a physical system with real-world data feedback to mitigate the complexities of predicting the performance of the system – and is useful in current and subsequent lifecycle phases of the system that it refers to (Qi et al. 2019) (Bilberg & Malik 2019). The advantages of having such a virtual twin of a physical system is to act as a forward-run to predict the future, identify any undesirables and suggest appropriate measures to avoid the undesirables (Grieves & Vickers 2017). It starts at the concept development phase of a system and continues to support decision making throughout development, operation, and maintenance. The concept of DT, besides other knowledge domains, has been widely researched for manufacturing applications (Schleich et al. 2017) (Haag & Anderl 2018) (Qi et al. 2019).

For digital twins of manufacturing systems, simulation capability is defined as an important, if not integral, aspect of a digital twin (Hehenberger & Bradley 2016) (Morandotti & Pelosi 2018) (Kuts et al. 2019). The eight dimension model of a digital twin described by (Stark et al. 2019) places the simulation capability at fifth dimension of achieving a system's digital twin.

The art of digitally modelling, analysing, and validating a manufacturing system has evolved from standalone 2D drawing to data intelligent digital twins. There are several types and forms of virtual simulation models (Mourtzis et al. 2014) that can be used to develop a digital twin of a product and its production system (see Figure 1).

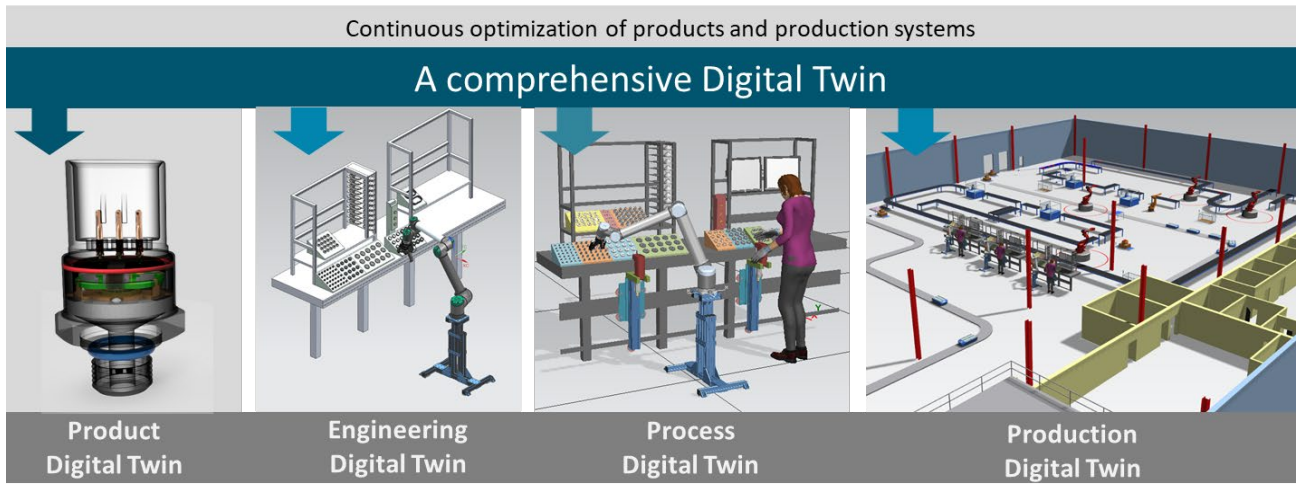

Figure 1. A comprehensive view of a digital twins.

The exactitude of the virtual models defines the accuracy of predicted behavior obtained from the experimentation on the digital twin. Furthermore, the efforts required to build these models need to be reduced and be more reconfigurable (parametric modeling).

The PLM packages offered by major companies such as Siemens, PTC, Dassault Systems, SAP and Oracle offer solutions for factory planning, design, and deployment (Tolio et al. 2013). But the lack of frameworks often makes it difficult to achieve a model with accuracy, short lead time and with ability to be reconfigured. This can also be described as value to cost ratio of building a digital twin which becomes difficult to justify if these attributes are not available.

This paper discusses the simulation domain of a digital twin for a manufacturing system. This paper presents a generic framework to model a production system integrating different forms of simulations at various levels of details that can subsequently be integrated with a digital twin.

The presented study and the DT simulation framework can be useful in situations:

- When evaluating a new design / strategy with or without owning the physical system
- When developing a virtual simulation of an already operational / legacy manufacturing system
- To investigate behaviour of a sub-system (e.g. machine, segment etc.) of a larger system (factory) with or without investigating its relationship with the rest of the system components

## 2. Background literature

### 2.1. Description of an adaptable manufacturing system

Manufacturing, as a technological process, is the application of physical and/or chemical processes to transform the properties, geometry or shape of a given starting material (Wiendahl et al. 2007) (Groover 2007). Quite often, it also involves a sequential integration or joining of components to achieve assembled products (Nof et al. 2012). An organized collection of equipment, people, and procedures to achieve these manufacturing operations is referred to as a manufacturing system (Groover 2007). While the responsiveness of a manufacturing system can be described by its ability to adapt to product variations and its capacity to adjust to fluctuations in demand (Koren & Shpitalni 2010).

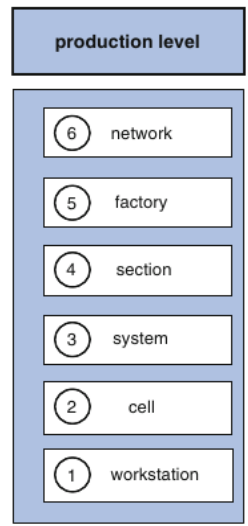

Figure 2. Levels of a factory (Wiendahl et al. 2007).

To simplify the study of a manufacturing system it may be classified into segments or levels. A classification of a manufacturing system was presented by (Wiendahl et al. 2015) as factory levels. The smallest level is a workstation while a cell comprises several workstations. The higher level refers to a factory (see Figure 2). Each level may have its own structure, composition, complexity, and relationship with upstream and downstream activities. The level of automation would also be different in each level. These levels have been discussed by (Wiendahl et al. 2007).

### 2.2. Simulations in manufacturing design and planning

Modern manufacturing systems present a mutual causality between complexity and desire for their predictability. In other words, the complexity of manufacturing systems is growing, and it is becoming difficult to predict a system's performance (Malik & Brem 2020). However, the need of predictability is becoming increasingly important to adapt the system to market variations.

A way to study the behavior of complex systems is through quantitative modeling. Quantitative models are often in the form of mathematical models (Banks & Carson II 1986). However, when mathematical models are used to study real-world systems, many times the increased complexity makes it difficult to achieve an analytical solution. In such situations, simulation modeling can be helpful (Figure 3). Simulation software translates mathematical equations into less abstract symbols (Misselhorn 2015), with a higher intuition and less mathematical representation yet rigorously based on mathematical models (Misselhorn 2015). Simulation models (as computer supported calculations) project past or future outcomes of the process, system or event. To conclude, computer simulation models are based on mathematics and are therefore mathematical models, yet, simulations are more suited to situations where analytical solution is difficult to achieve due to increased complexity.

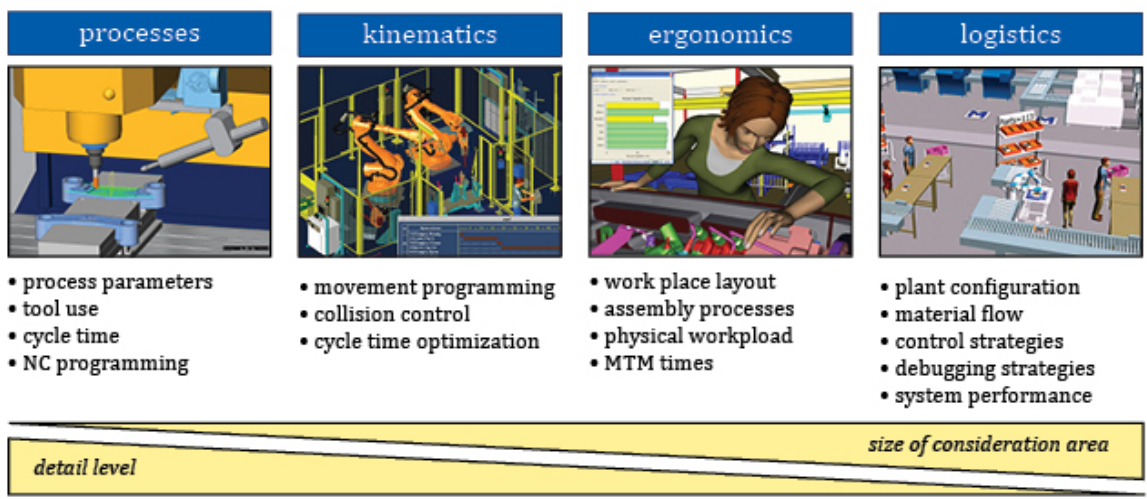

Figure 3. Different forms of simulation in production systems; adapted from **(Wiendahl et al. 2015)**.

Simulations, in a broader context, can be developed with physical objects, virtual objects or a combination of both, but a computer simulation is using of a computer software for building and coding an operating and quantifiable model of real-world process, system, or events (Law et al. 2000) (Davis et al. 2007), whose variables can be manipulated (Berends & Romme 1999) and the resulting output is available immediately.

Once a simulation model is developed, the time of a simulation can be fast forwarded or reversed to observe the effect of manipulation of variables to reach a specific instance of time and seeing the artificial past or future (similar to time-machine concept). The accuracy of a model comparing to the real world is important to achieve reliable results. Although this paper is focused on computer based digital simulations, yet the combination of both the physical and virtual simulations are also discussed.

## 2.3. From simulation to digital twin

CAD models generated during development of a new production system contain maximum details of the desired performance of the system (Bilberg & Malik 2019). Efforts have been made to extend these models to various types of simulations thus extracting as much information as possible before making any investment. Furthermore, the increasing need of customization requires that manufacturing systems are continuously evolving and rapidly reconfigured (Masood & Sonntag 2020). Hence the changeover times are getting short while it is important that the reconfiguration strategies are validated before implementation into the real world (Thomas et al. 2016) (Leng et al. 2020).

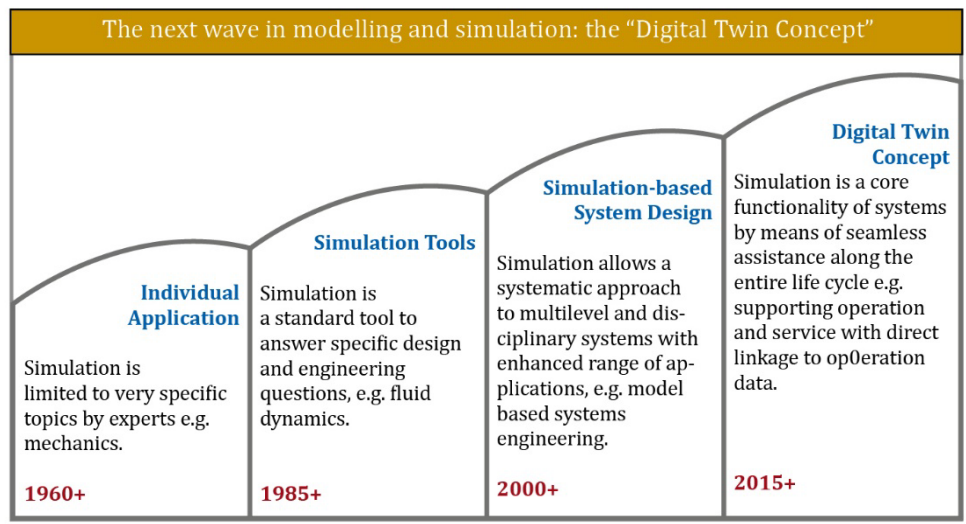

Figure 4. The evolution of simulation technology, adapted from (Hehenberger & Bradley 2016).

Hardware in the loop simulations (Harrison III & Proctor 2015) can speed up the changeovers (Malik et al. 2020). Furthermore, the recent years have seen considerable development (see Figure 4) in using simulation models along the lifecycle of a system and are referred to as a digital twin (Malik & Brem 2020) with seamless data connectivity between the real system and the simulation model (see Figure 5). Thus, the real feedback from the real system acts as input data for the simulation and enables a more realistic and faster performance optimization (Qi et al. 2019).

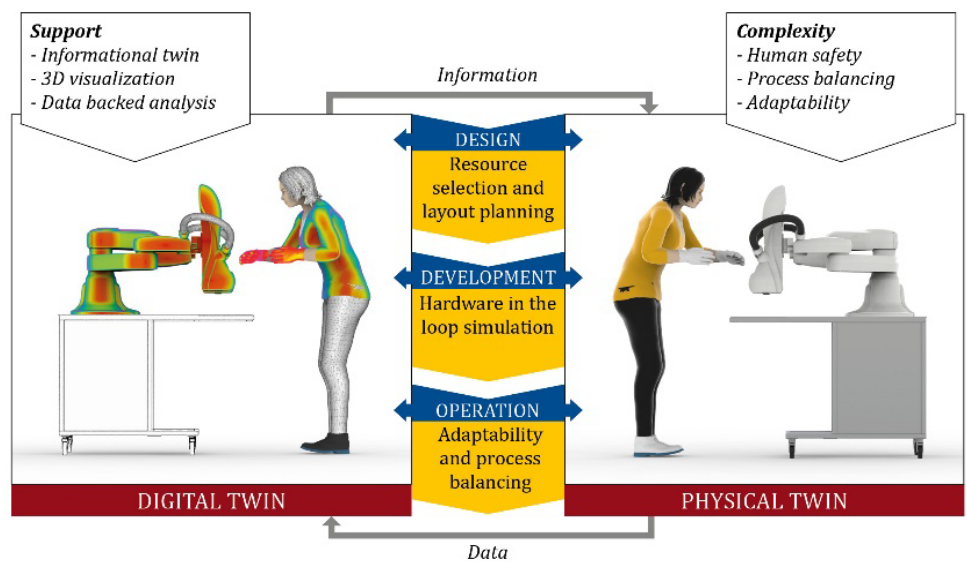

Figure 5. Digital twins in system design (Malik & Brem 2020).

## 3. VMDT framework for modeling of a production system

The simulation modelling of a manufacturing system is affected by complexity, detail level and the size of the considered area. It is also influenced by the problem for which the digital twin needs to be created as it will define the required details of the simulation model. A generic framework for simulation modelling of a manufacturing system is shown in Figure 6. The model illustrates the flow of a detailed simulation analysis as well as its integration with the hardware to build HiL (hardware in the loop) systems or a digital twin.

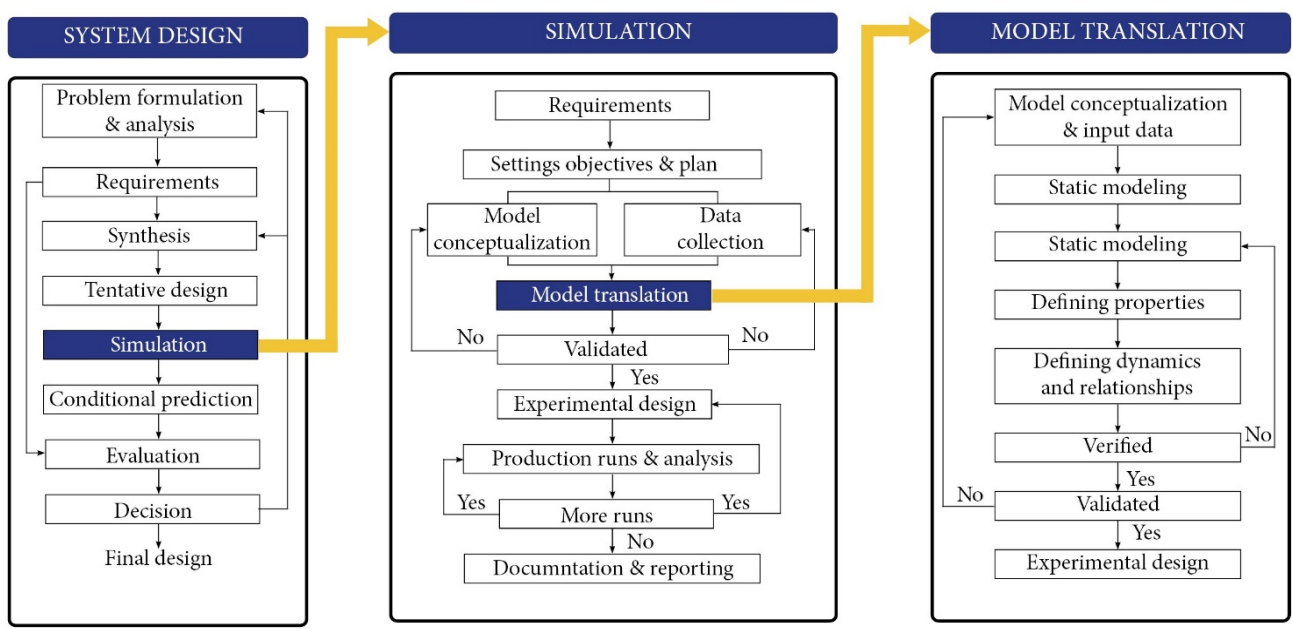

Figure 6. VMDT framework for virtual modeling for a digital twin.

### 3.1. Virtual modeling

The modeling is done in three steps i.e. static modeling, defining dynamic relationships, defining properties and attributes.

### 3.1.1. Static modeling

Static modeling is primarily the creation of static visualization of the system under observation. It starts by creating CAD objects and placing them in an approximate order thus representing the manufacturing system (Figure 7). These models can be built in Computer Aided Designing (CAD) software such as CATIA, NX, Creo, SolidWorks etc. Over the years, different CAD formats have been introduced to enable data exchange from one CAD system to another. Most common 3D data exchange formats are STEP, and IGES. It is becoming increasingly important that 3D data of machines and resources is made available by the OEMs.

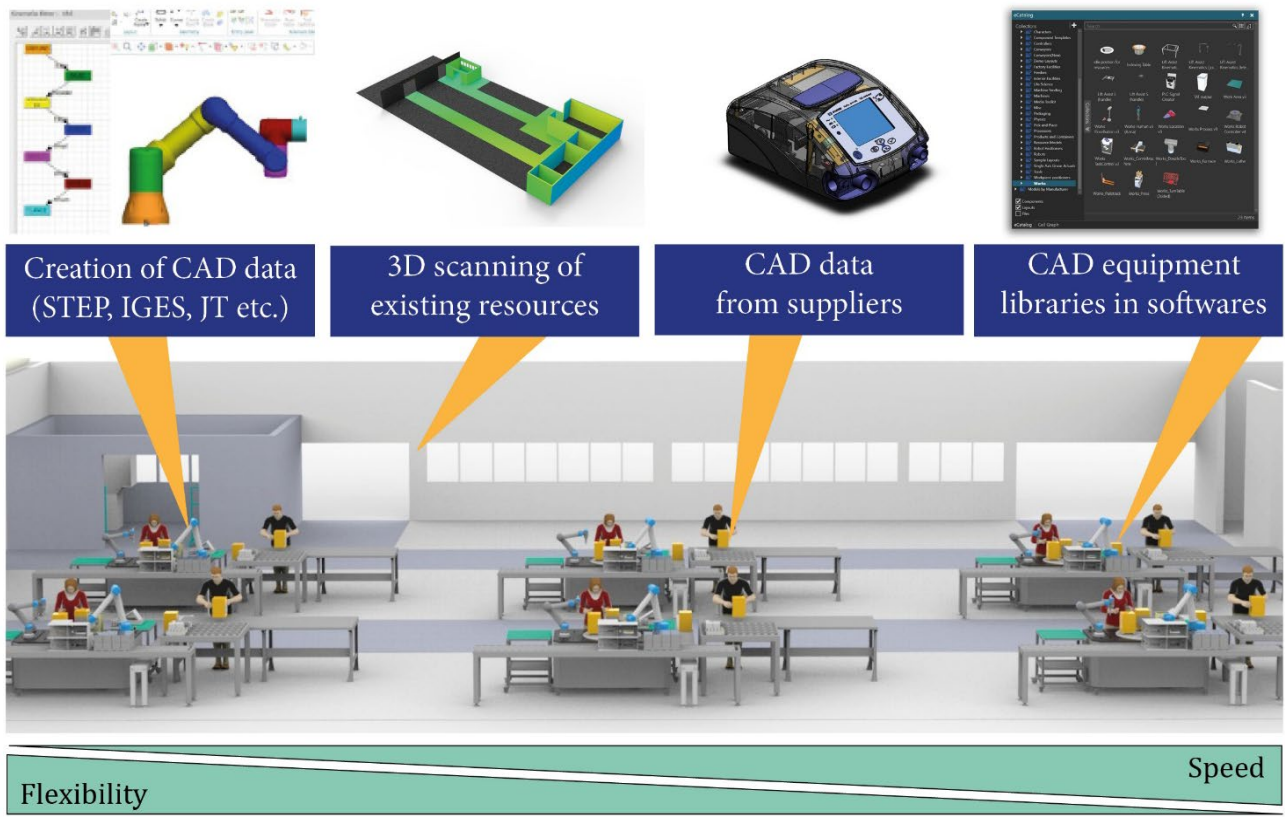

Figure7. Static modeling of production systems.

It can be time consuming and has a prerequisite that correct level of details is identified to avoid wasteful efforts in creating details that are not required. It is also important that depending upon the complexity of the system, correct data format is used. For example, STEP and IGES are often heavy CAD formats while JT is lighter. Surface models can be helpful if study of the physical forces is not required.

It is time consuming to build high fidelity virtual models of a complex production system. Jain (Jain et al. 2015) proposed that joint efforts are needed to simplify this step. Most manufacturers of factory equipment are now offering CAD 3D data of their products at various detail levels in standardized CAD data exchange formats (e.g. STEP, OBJ, JT etc.). The most common data format is STEP which is usable in most modeling software.

Another development is the availability and integration of standard libraries (Fig. 6) in the modelling software. Factory design packages are now being offered by various venders with built in or cloud-based libraries of standard production resources. These resources, once used in the virtual model, can be exported to simulation software.

As being said, a production system can have various levels of details starting from a workstation to production floor, and site. Resources are combined together in desired quantity and order to achieve the desired production. For virtual experimentation, it is a first step to have computer-based models of these resources. The CAD models of the components and assemblies are then combined together and placed at appropriate locations in the virtual environment thus creating a system.  The placement of the resources or the layout can be optimized using features offered by CAD tools.

### 3.1.2. Defining properties and attributes

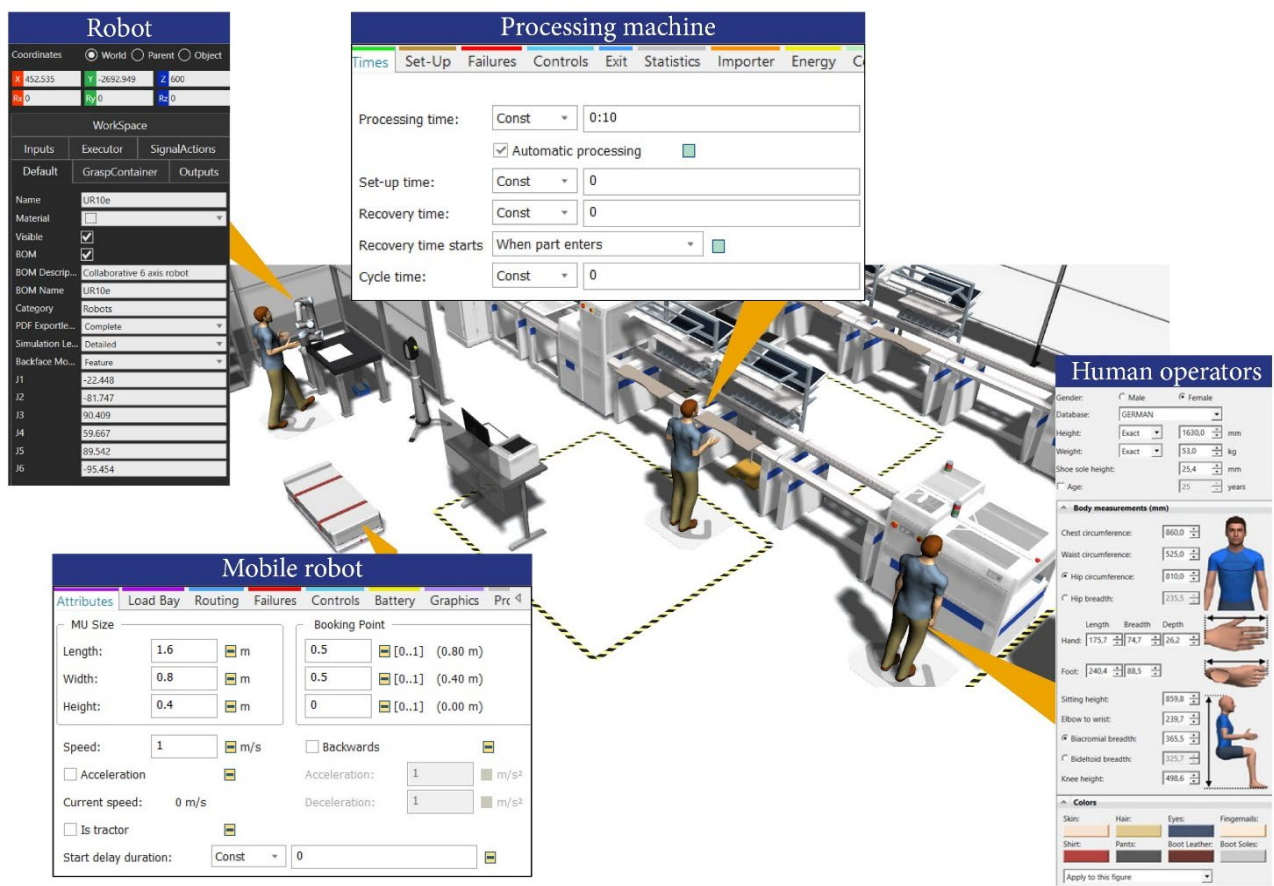

Figure 8. Defining properties of production resources in virtual modeling.

After static CAD models of production resources have been created the models can be imported into a simulation environment where kinematic simulations (often time-based continuous simulation) or material flow simulation (discrete event simulation) can be performed. Each resource is placed at appropriate locations defining an initial layout and the next step is to define relevant properties, attributes and physical characteristics for each active production resource. The workspace may also include passive resources that do not affect the production output and maybe ignored for simulation.

The Figure 8 shows an example of a simulation model of a production space that includes a human operator, a mobile robot, a machining center, and some passive resources. The properties of each of the active resource are also shown in the figure.

### 3.1.3. Dynamic relationships

The next step is to define the dynamic behavior of the virtual production resources. It includes defining relationships between entities of a static CAD model. In case of a machine it may include the relationship and constraints between different components. Human resources (Figure 9) required in the production system can also be added. Material flow constraints are added (Figure 10).

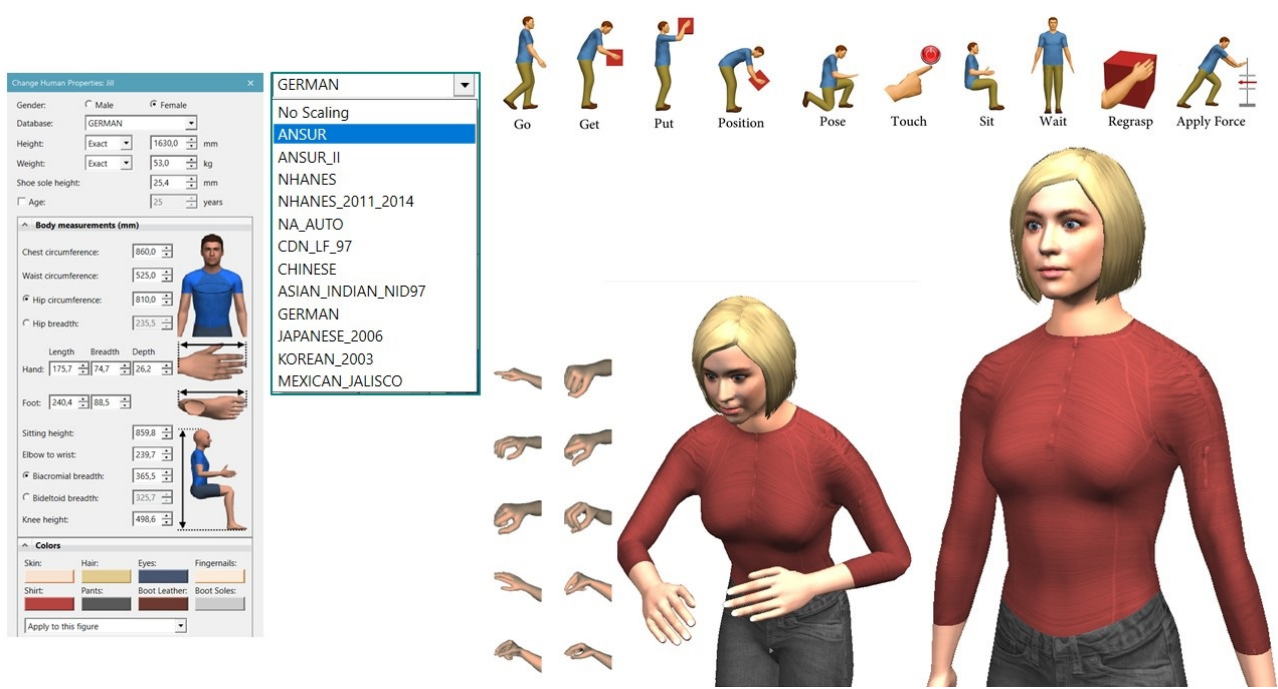

Figure 9. Digital human modeling.

### 3.2. Simulation runs

A real world doesn't work with uninterrupted ideal situations. Therefore, variables need to be introduced for realistic analysis. (Heilala & Voho 2001) described DES the best method to evaluate variability of reconfigurable assembly cells (Figure 11). The model developed in previous chapter with continuous simulation is imported into a DES software. The software that can be used in such study are Visual Components, Tecnomatix Plant Simulation, Simio, FlexSim, AnyLogic, ProModel etc.

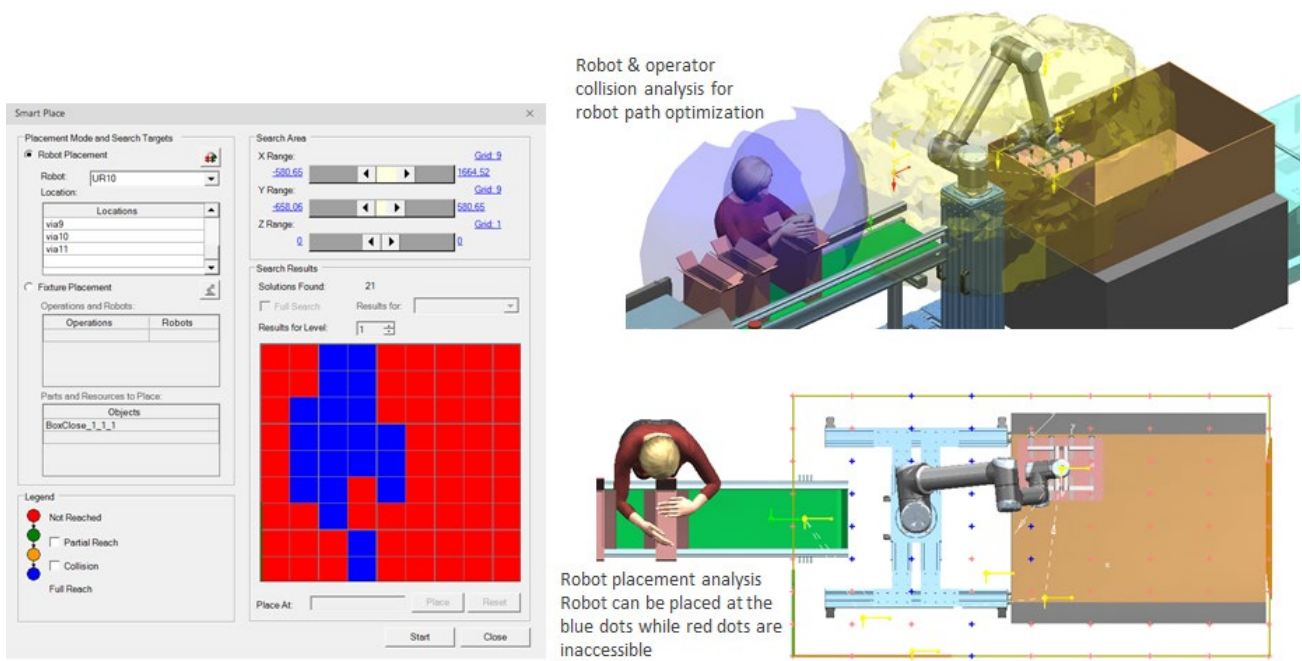

Figure 10. Continuous simulations in design of production systems.

Continuous simulation helps to calculate accurate estimations for the cycle times for each activity depending upon e.g. robot trajectories, human posture, and time to transfer a product from one station to another. It not only generates a realistic simulation of the process but also calculates average cycle times, robot cycle times given optimized robot trajectories, human collisions with the robots and safety analysis for implemented safety shielding.

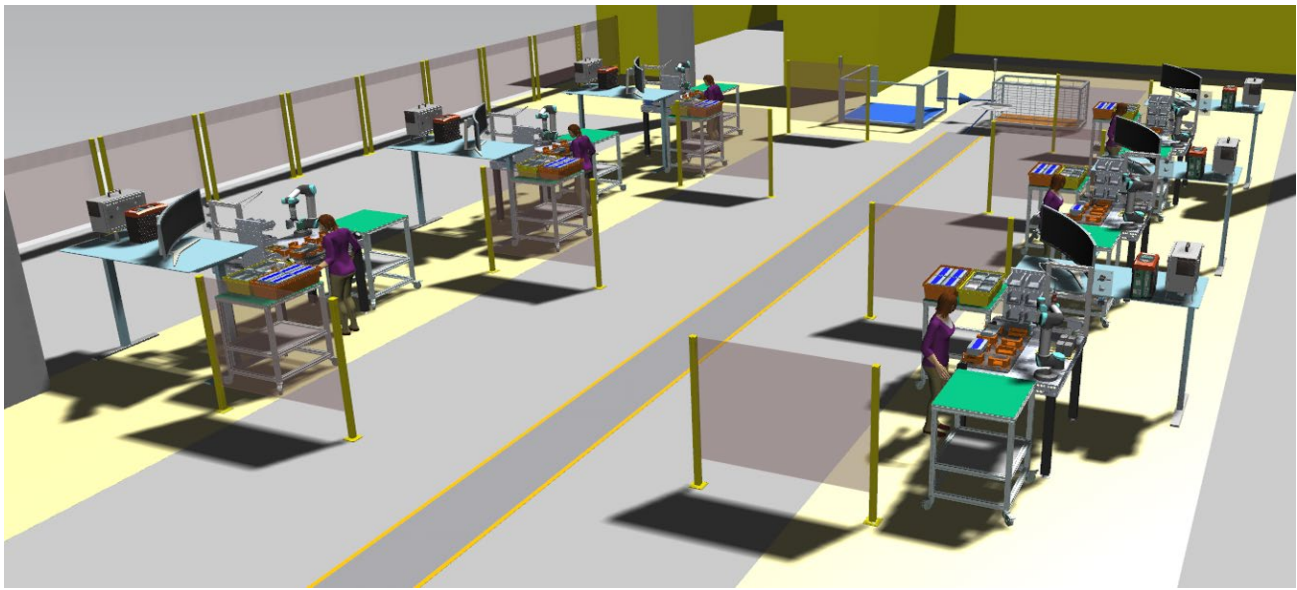

Figure 11. Discrete event simulations in planning and optimizing production scenarios.

This type of modelling can be performed in commercial platforms such as Tecnomatix Process Simulate, DELMIA, Visual Components etc. The modular approach (Guo et al. 2019) is utilized for simulation modelling in this phase as well and classes (objects) are created with given parameters for the equipment.

## 4. Commercial aspects of simulation modeling

The need and importance of simulation modeling is well acknowledged. But the industrial application particularly for SMEs is quite limited. The challenges that are often regarded a hindrance in industrial usability of simulations are lack of knowledge, lack of skilled people, cost of software, and retention of trained personnel. Some of the challenges such as long time spent on building the simulation models maybe reduced by defining the right type of simulation study and breaking it down to different levels. A theoretical model (Figure 12) in this regard is presented in the figure which acts as a stage gate model. The objectives of defining a simulation study can be decomposed into several phases and each can be approached depending upon the need.

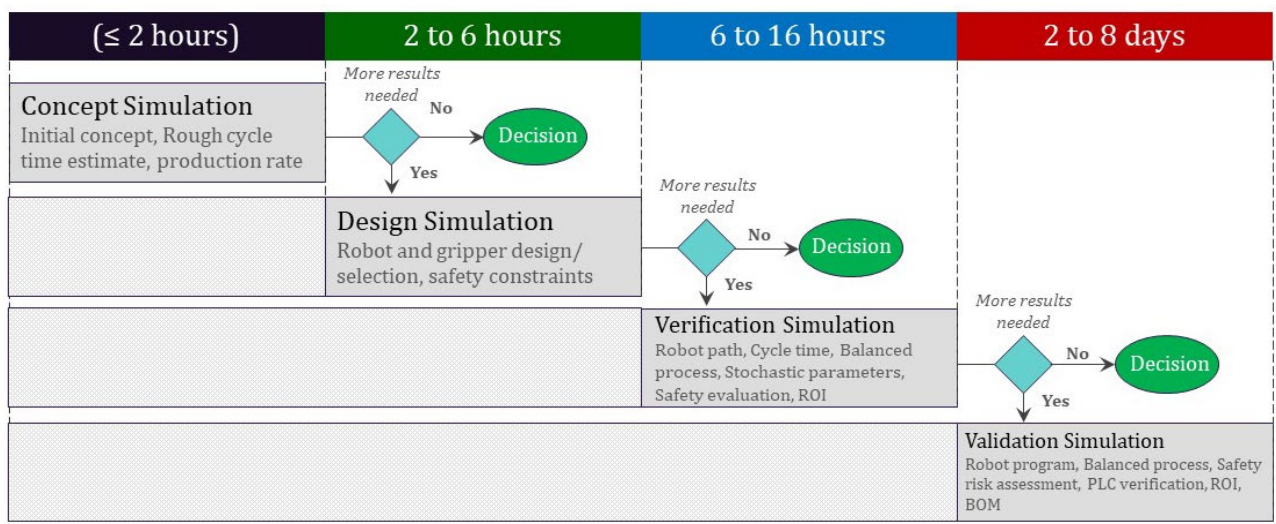

Figure 12. Using simulations in manufacturing industry.

## 5. User interaction with the virtual models

Humans or users can interact with the virtual models using virtual screen displays. The integration of physical and virtual objects can be enabled to have fluid interaction of the naive users with the simulation (Figure 13). A growing interest is there to interact with the virtual models in immersive environments. The immersive environments let the user to interact with the equipment by virtually entering the scene (Figure 14). More intuition increases productivity and effectiveness of simulation modeling. For example, enabling the use of vocal commands in addition to mouse and keyboard interaction between software and user results in increased productivity (Feeman et al. 2018). Augmented reality is another way to interact and evaluate the virtual models. AR holograms (Fig. 10) are embedded with the real world to visualize the results. A possible result is to see the placement of a virtual machine in a room.

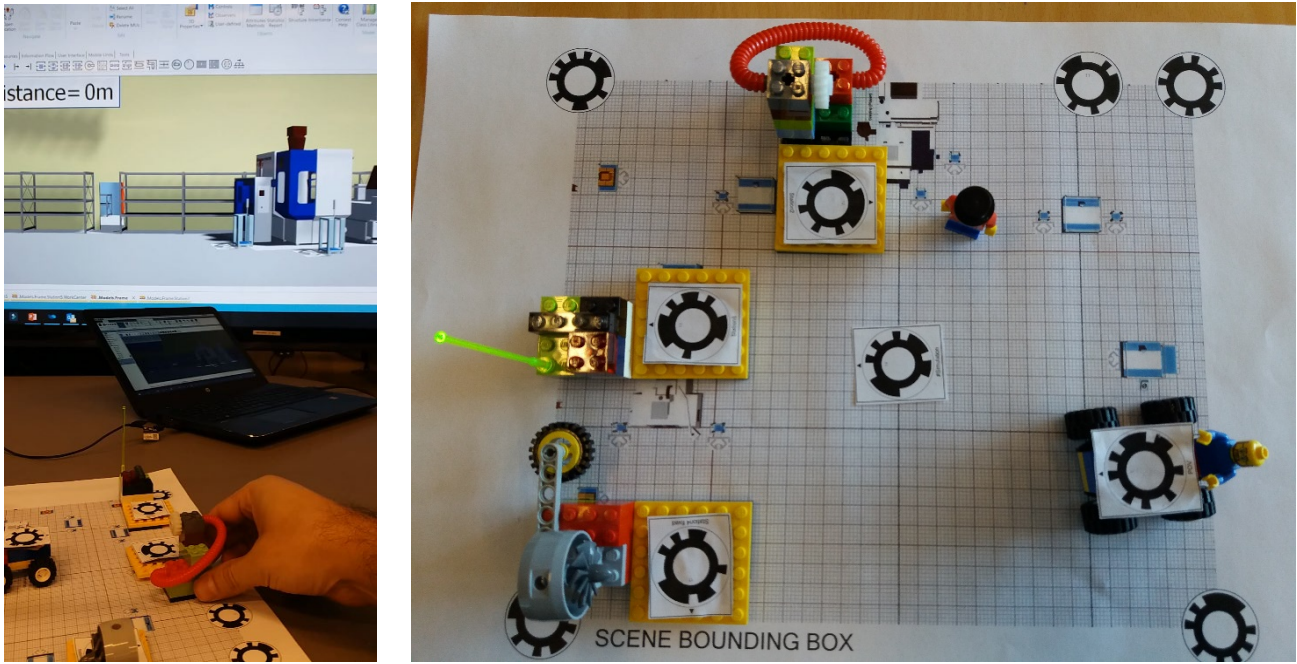

Figure 13. Fluid interaction between the end users and the simulation.

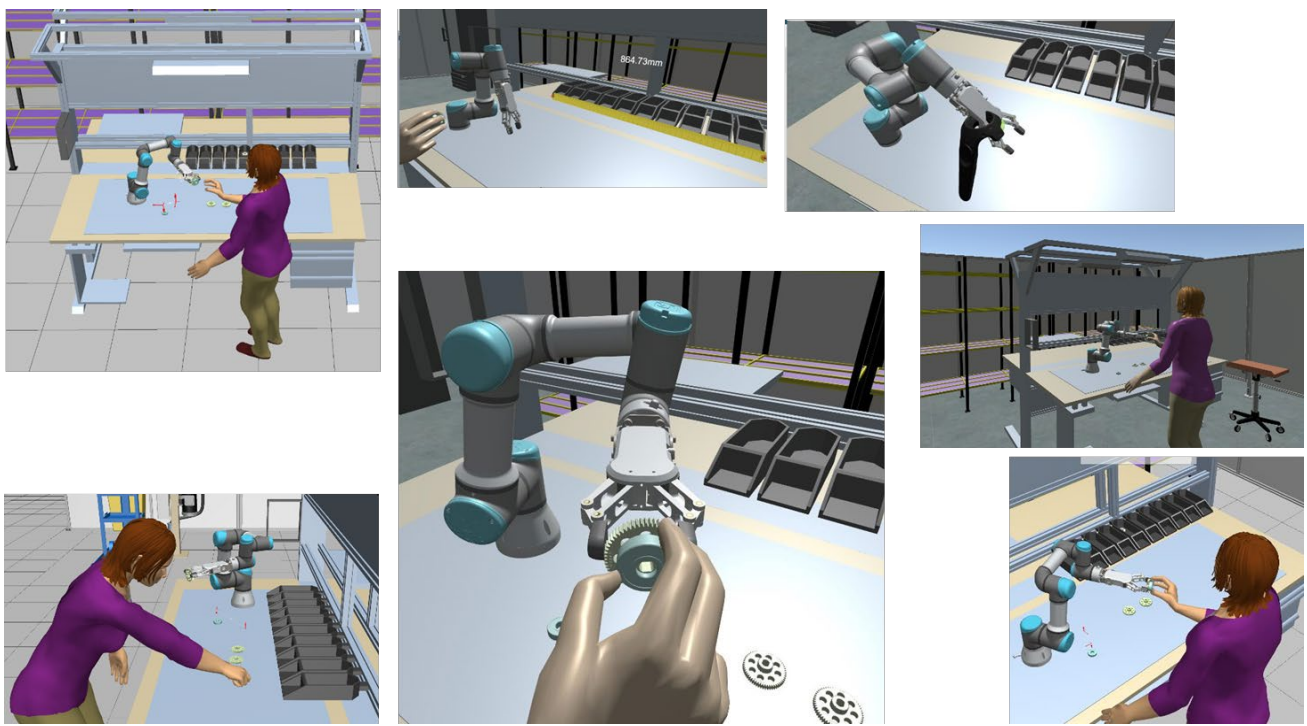

Figure 14. Immersive environment to interact with the simulation objects.

## 6. Interfacing simulation with physical system

The integration of simulation model with the physical counterparts can have multiple advantages (Figure 15). An example of it is industrial robotic. A great deal in robotic applications is associated with the need of expertise and effort to program the robots (Rossano et al. 2013). Even in case of cobots – with an easier programming interface - the programming is still a time consuming and tedious task for industrial applications. With the use of a DT, the robot program is intuitively generated in a simulation environment (Figure 16). Once the desired operation is tested virtually with defined robot trajectories and transition logics the robot program is transferred to the connected cobot that starts working as the robot in the digital twin (offline). An online connection between physical and virtual robot can dynamically transmit any movement made in physical space to the robot in the virtual space and vice versa (Fig. 11). Thus, avoiding any need to have any additional programing. Since the simulation has robot paths in a robot understandable language, the robot program can be transferred back to the simulation to make simulation run according to the physical robot.

**Sensor interfaced with the simulation**
1: Arrival of boxes
2: Conveyer blocked
3: Box filled
4: Boxes ready to be picked
5: Robot parameters
6: Box filled
7: Mobile robot ready

**Possible advantages**
- Comparison of simulation data with real data
- More informed production system
- Data driven simulations to make correct estimates
- Data driven decision making for resource allocation
- Realtime future monitoring and advise

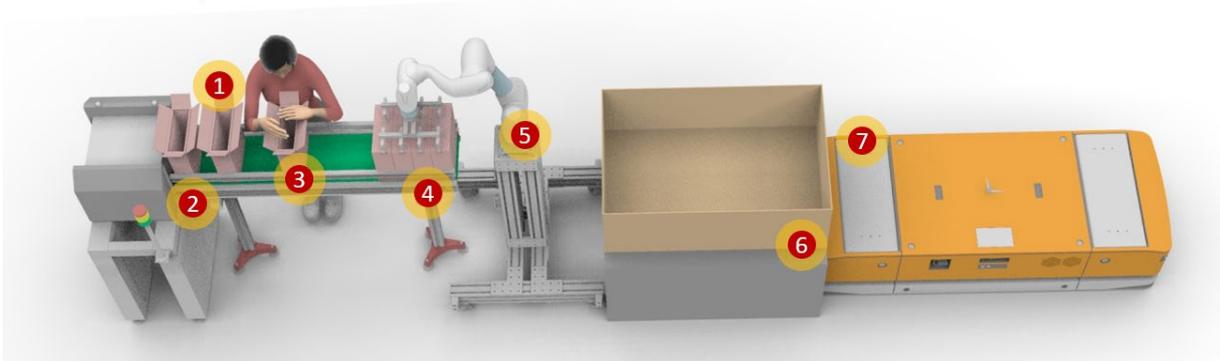

Figure 15. Interfacing the sensor data with a simulation to build a real-time digital twin.

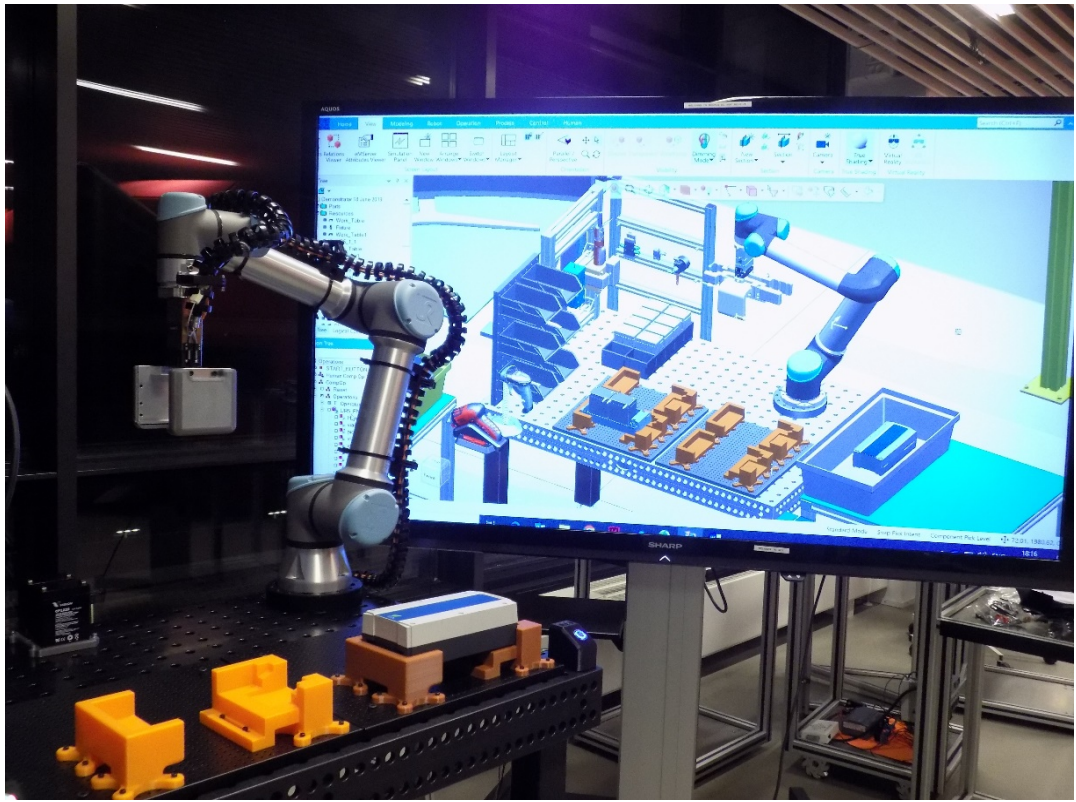

Figure 16. Interfacing robot feedback with its digital twin.

## 7. Discussion and conclusion

With growing demand of customization, challenges of globalization, and the penetration of information and communication technologies, the future factories are expected to be adaptable, responsive, and connected. This vision is often referred to as Industry 4.0 or the fourth industrial revolution. The infusion of enabling technologies the future manufacturing systems are seen as complex biomechatronic systems.

Digital twin has emerged as the next wave in the simulation technology (Hehenberger & Bradley 2016) to answer the growing complexities of manufacturing systems. Although the predictions of a system's performance can also be developed using mathematical computational models or simulations (Narayanan et al. 1998). But a digital twin, as a data connected digital shadow of a physical system, can help deal with complexity that grows along the lifecycle of a production system. Modeling and enabling a DT is useful if the system that it refers to is:

- of complex nature
- continuously evolving
- difficult to predict with conventional simulations

Digital twin is the concept to build a virtual model as a 4-dimensional model of a physical system. In factory operations, the increasing integration of internet, PLCs and computers is making the systems complex and the frequent changes require that the system is predictable. DT aims to solve the question by building a virtual model often with a 3D representation and with a 4th dimension of time or duration.

A successful implementation of a digital twin of a factory can integrate the whole lifecycle of the product in a virtual or digital world.

The economic justification of building a digital twin is easier said than done. However, the information obtained from a digital twin model is a replacement for wasted physical resources i.e. time, energy and material (Grieves & Vickers 2017). Therefore, if the cost of information is less than wasting the resources then information must be traded off.

Simulation-based digital prototyping is a well-grounded method in research and practice to support decision making in design and reconfiguration of manufacturing activities (Flores-Garcia et al. 2015), yet the reduction of development time through the use of CAD tools in manufacturing systems' development has not been parallel to the advancement in product development (ElMaraghy 2005). Digital simulations make it possible to make insights into complex manufacturing systems to form and validate optimizations before translating them into the real world (Mourtzis 2019) (Uhlemann et al. 2017). Reduction in development time and cost is the primary motive to make virtual experimentation and validation (Azab et al. 2012). As the need for changeability and flexibility is increasing, the need for virtual experimentation is also gaining more attention (Sølund 2017).

Robotics simulations are used to plan, predict, and safely test what if scenarios when planning a robot based system (Schluse & Rossmann 2016). Similarly, the digital human modeling, through integrating a computer-rendered avatar, can examine ergonomic viability of human actions (Chaffin 2008) (Poirson & Delangle 2013) (Schuh et al. 2015).